%% file: FinalManuscript.tex
\documentclass[fleqn,usenatbib]{mnras}

\usepackage{newtxtext,newtxmath}

\usepackage[T1]{fontenc}
\usepackage{ae,aecompl}

\usepackage{float}
\usepackage{hyperref}
\usepackage{color}
\usepackage{mathrsfs}
\usepackage{epsfig}
\usepackage{epstopdf}
\usepackage{multirow}
\usepackage{longtable}
\usepackage{lscape}
\usepackage[flushleft]{threeparttable}
\usepackage{caption}

\usepackage{graphicx}	
\usepackage{amsmath}	
\usepackage{amssymb}	

\newcommand{\mum}{${\rm \mu m}$}

\newcommand{\Lx}{$L_{X}$}

\newcommand{\z}{$z$}
\newcommand{\Mstar}{$M_\ast$}
\newcommand{\Msun}{$M_\odot$}
\newcommand{\ergps}{erg s$^{-1}$}

\newcommand{\Rms}{$R_{\rm MS}$}

\title[A difference in the SFR of low versus high \Lx\ AGNs]{Inferring a difference in the star-forming properties of lower versus higher X-ray luminosity AGNs.}

\author[E.~Bernhard et al.]{
E. Bernhard,$^{1}$\thanks{E-mail: e.p.bernhard@sheffield.ac.uk}
L. P. Grimmett,$^{1}$
J.R. Mullaney,$^{1}$
E. Daddi,$^{2}$
C. Tadhunter,$^{1}$
and S. Jin$^{2,3}$
\\
$^{1}$Department of Physics $\&$ Astronomy, University of Sheffield, Sheffield S3 7RH, UK\\
$^{2}$ CEA, IRFU, DAp, AIM, Universit\'e Paris-Saclay, Universit\'e Paris Diderot, Sorbonne Paris Cit\'e, CNRS, F-91191 Gif-sur-Yvette, France\\
$^{3}$School of Astronomy and Space Science, Nanjing University, Nanjing 210093, China
}

\date{Accepted XXX. Received YYY; in original form ZZZ}

\pubyear{2018}

\begin{document}
\label{firstpage}
\pagerange{\pageref{firstpage}--\pageref{lastpage}}
\maketitle

\begin{abstract}
We explore the distribution of \Rms$\equiv$SFR/SFR$_{\rm MS}$ (where SFR$_{\rm MS}$ is the star formation rate of ``Main Sequence'' star-forming galaxies) for AGN hosts at \z=1. We split our sample into two bins of X-ray luminosity divided at \Lx=2$\times 10^{43}$\ergps\ to investigate whether the \Rms\ distribution changes as a function of AGN power. Our main results suggest that, when the \Rms\ distribution of AGN hosts is modelled as a log-normal distribution (i.e. the same shape as that of MS galaxies), galaxies hosting more powerful X-ray AGNs (i.e. \Lx$>2\times 10^{43}$\ergps) display a narrower \Rms\ distribution that is shifted to higher values compared to their lower \Lx\ counterparts. In addition, we find that more powerful X-ray AGNs have SFRs that are more consistent with that of MS galaxies compared to lower \Lx\ AGNs. Despite this, the mean SFRs (as opposed to \Rms ) measured from these distributions are consistent with the previously observed flat relationship between SFR and \Lx. Our results suggest that the typical star-forming properties of AGN hosts change with \Lx, and that more powerful AGNs typically reside in more MS-like star-forming galaxies compared to lower \Lx\ AGNs.
\end{abstract}




\section{Introduction}

It is now recognised that the activity caused by the growth of supermassive black-holes (SMBHs) at the centre of galaxies (observed as Active Galactic Nuclei; AGNs) has played a major role in shaping today's galaxies \citep[e.g.][]{Gebhardt2000, King2003}. However, although there are multiple lines of empirical evidence showing that SMBH growth is {\it on average} related to the growth of their host galaxy via star-formation \citep[see][for a review]{Harrison2017}, there is no clear consensus on the physical mechanisms (should it be e.g. feedback or common fuel--triggering mechanism) that generate these trends between average SMBH and galaxy growth.

To better understand the impact of SMBH growth in galaxy evolution, one can measure the star formation rate (SFR) of a large sample of AGN hosts at multiple epochs. Using {\sc Herschel}\footnote{{\sc Herschel} is an {\sc ESA} space observatory with science instruments provided by European-led Principal Investigator consortia and with important participation from NASA.}, which provides an unprecedented view of the galaxy star-formation at far-infrared (FIR) wavelengths, recent studies have found that (1) there is no relationship between mean SFR and X-ray luminosity (\Lx, a proxy for AGN power, e.g. \citealt{Stanley2015, Lanzuisi2017, Stanley2017}) and (2) that the mean AGN host SFR is broadly consistent with that of normal star-forming galaxies \citep[e.g.][]{Mullaney2012b, Stanley2015} for which the SFR is correlated to the stellar mass via the Main Sequence (MS; e.g. \citealt{Schreiber2015}, hereafter S15). However, although {\sc Herschel} provides the deepest view of SFRs at FIR wavelengths, a large fraction of AGN hosts (typically more than 50 percent) are not individually detected, meaning most studies rely on method such as stacking to obtain averages. As these averages can potentially be dominated by bright outliers, the empirical mean SFR of AGNs might not be representative of the ``typical'' SFRs of the full AGN sample, increasing the complexity of investigating the AGN-galaxy connection \citep[e.g.][]{Mullaney2015, Scholtz2018}.

Instead of relying on mean SFRs, \citeauthor{Mullaney2015} (\citeyear{Mullaney2015}; hereafter M15) have measured the full distribution of SFRs relative to that of the MS (\Rms$\equiv$SFR/SFR$_{\rm MS}$) for AGN hosts out to \z$\sim$4 using a combination of {\sc Herschel} and {\sc ALMA} observations. They found that the mean-average of the \Rms\ distribution is consistent with that of the MS, yet the mode (i.e. the most common value) lies below that of the MS. This is a consequence of the mean being enhanced by bright outliers, leading to a biased picture. Furthermore, they report a \Rms\ distribution for AGN hosts twice as broad as that of MS star-forming galaxies, demonstrating that the star-forming properties of AGN hosts are more diverse than that of MS galaxies. However, interestingly, they do not find any evidence of a significant evolution of the \Rms\ distribution with redshift. More recently, \cite{Scholtz2018} have measured the distribution of specific SFRs (SFRs relative to stellar masses; sSFRs) for massive (i.e. \Mstar$>2\times10^{10}$\Msun) galaxies hosting bright (i.e. \Lx$>10^{43}$\ergps) AGNs in a large range of redshifts (i.e. 1.5$<$\z$<$3.2), and find good agreement with that of simulated galaxies taken from the {\sc EAGLE} simulation \citep[see][and references therein]{Scholtz2018}. They report no differences in the distribution of sSFR with \Lx\ for AGNs with \Lx$>10^{43}$\ergps. As demonstrated in \cite{Scholtz2018}, while the sSFR distribution hints important information regarding the connection between AGN and their host galaxies, it lacks of context in terms of the MS of star-forming galaxies. Instead, the \Rms\ distribution and how it changes with \Lx\ provides a better insight of the star-forming properties of AGN hosts within this context of the MS of star-forming galaxies.

In this work, we propose to expand upon M15 and measure whether the \Rms\ distribution changes with \Lx. As there is no apparent evidence of the \Rms\ distribution evolving with redshift (M15), we focus on AGNs at \z=1 (i.e. close to the peak of activity for both SMBH accretion and SFR; \citealt{Aird2015}). We describe our sample selection and sample properties in \S\,\ref{sec:sample}. We present our analysis in \S\,\ref{sec:analysis}, the results of which are shown in \S\,\ref{sec:results}. The implications of our results are discussed in \S\,\ref{sec:discussion}, and we conclude in \S\,\ref{sec:conclusion}. Throughout, we adopt a {\sc WMAP--7} year cosmology \citep{Larson2011} and a \cite{Chabrier2003} initial mass function (IMF) when calculating stellar masses and SFRs.

\section{Sample selection and properties}
\label{sec:sample}

\begin{figure}
 \centering
 \includegraphics[width=0.4\textwidth]{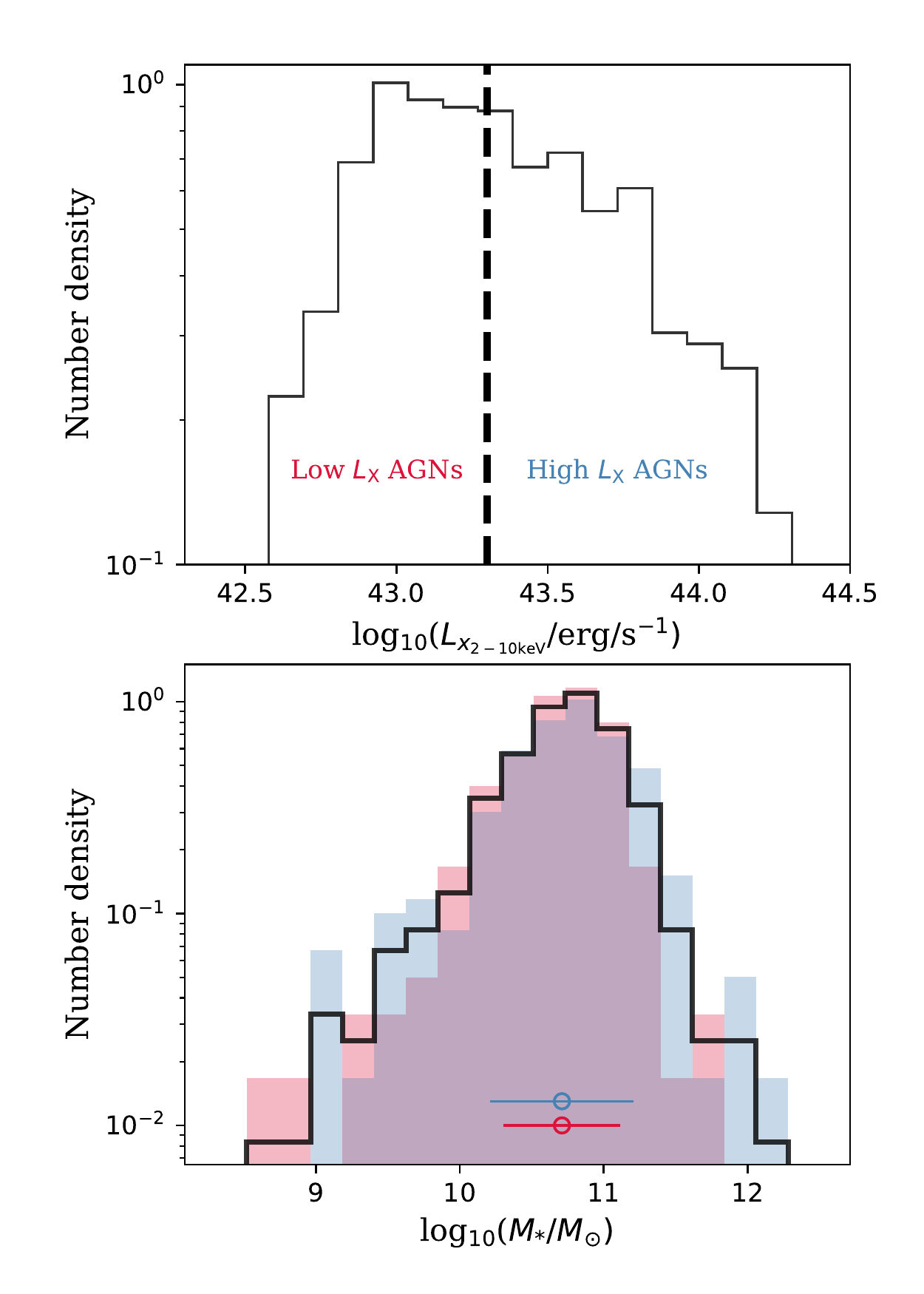}
 \caption{{\it Top}: The normalised distribution of intrinsic X-ray luminosities for our full sample of AGNs. The dashed vertical line shows our limit for lower and higher \Lx\ AGNs. {\it Bottom}: The normalised distribution of stellar masses for our full sample of AGNs (black line), our low \Lx\ sample (red histogram), and our high \Lx\ sample (blue histogram). The red and blue circles indicate the positions of the median masses for the low and high \Lx\ sample, respectively, along with their 1$\sigma$ uncertainties. \label{fig:distribs}}
\end{figure}

Our sample of X-ray sources is from the catalogue of \cite{Marchesi2016} that provides absorption-corrected 2-10~keV \Lx\ for AGNs in the {\sc COSMOS} field. We only retain sources that have 0.8$<$\z$<$1.2 to probe AGNs around \z=1 and minimise the effect of the SFR evolution with redshift (S15). We find that 776 AGNs in \cite{Marchesi2016} satisfy this requirement ($\sim75$~percent of which have spectroscopic redshifts), among which 664 are also covered by {\sc Herschel} observations. We further remove 123 X-ray sources out of the 664 that have upper-limits on their intrinsic X-ray luminosities as they mostly affect the lower luminosity range (i.e. \Lx$\sim10^{42\--43}$\ergps) and could be associated with star-formation activity. We show in the top panel of Fig.\,\ref{fig:distribs} the distribution of \Lx\ for our full sample of 541 AGNs. We match these X-ray AGNs to the catalogue of \cite{Jin2018} that contains ``super-deblended'' IR photometry (i.e. at 24\mum, 100\mum, 160\mum, 250\mum, 350\mum\, and 500\mum) for the {\sc COSMOS} field measured using a method outlined in \cite{Liu2018}. Of these, we find that 100 (i.e. $\sim$18~percent) show no detection in any of these IR wavelengths. For each of these, we derive 3$\sigma$ upper limits\footnote{The 1$\sigma$ upper limit are estimated using the standard deviation of a flux distribution built by performing 100 times aperture photometry on randomly selected positions located within the full width at half maximum of the point spread function. The aperture corrections are publicly available at \url{http://www.mpe.mpg.de/resources/PEP/DR1_tarballs/readme_PEP_global.pdf}.} at 100\mum\ and 160\mum\ using the {\sc COSMOS} maps provided by the {\sc PACS} Evolutionary Probe team \citep{Lutz2011}. Our final sample of X-ray selected AGNs contains 541 sources with IR detections in at least one of the following bands: 24\mum\ (81 per~cent), 100\mum\ (21 per~cent), 160\mum\ (15 per~cent), 250\mum\ (35 per~cent), 350\mum\ (22 per~cent), and 500\mum\ (7 per~cent), or upper limits at 100\mum\ and 160\mum.

To measure SFRs, we use a similar approach than that of \cite{Bernhard2016} which employs multi-component IR spectral energy distribution (SED) fitting using {\sc DECOMPIR}\footnote{{\sc DECOMPIR} is publicly available at \url{https://sites.google.com/site/decompir/}} \citep{Mullaney2011}. Briefly, {\sc DECOMPIR} performs chi-square minimisation to select the best combination of one out of five templates for the host galaxy emission\footnote{See \cite{Mullaney2011} for a full description of the galaxy templates that are available in {\sc DECOMPIR}.} and an empirically derived AGN template (see \citealt{Bernhard2016} and references therein for details on the fitting approach). The IR luminosities arising from star formation are then derived from the fits (after removing the AGN contamination), and converted to SFRs using equation (4) in \cite{Kennicutt1998} adapted for a \cite{Chabrier2003} IMF. This fitting approach is applied to the 30~percent of our sources that have IR SEDs with at least three photometric points. For the remaining 70~percent of the sources we derive SFR upper limits by only fitting the host galaxy templates (i.e. ignoring AGN contamination) using {\sc DECOMPIR} when the AGN is only detected in two photometric bands, or by using the most common of {\sc DECOMPIR} templates (i.e. ``SB2'') found for our sample with multiple detections, and for which we choose the highest normalisation that does not over-predict any detected photometric points or upper limits at 100\mum\ and 160\mum. These are SFR upper limits since, should the IR be contaminated by AGN emission, it would decrease the contribution of the host to the IR luminosities, hence SFRs. As our aim is to measure the distribution of SFRs for AGNs relative to that of the MS, we also require host stellar masses. These are derived using {\sc CIGALE} \footnote{{\sc CIGALE} is publicly available at \url{https://cigale.lam.fr}} which performs a multi-component ultraviolet-to-IR SED fits accounting for AGN contamination \citep{Noll2009, Ciesla2015}, and shown in the bottom panel of Fig.\,\ref{fig:distribs}. Our stellar mass estimation is fully presented in Grimmet et al. (subm.). However, as the stellar masses in {\sc CIGALE} can be affected by contamination from unobscured AGNs \citep{Ciesla2015}, we also performed our analysis using only obscured AGNs (representing roughly 60~percent of our full sample). In doing so,  we find consistent results compared to using the full sample, suggesting that our results are robust to any biases in our mass estimation. The SFR of the MS is derived using equation (9) of S15 adapted for a \cite{Chabrier2003} IMF.

To explore how the distribution of the star-forming properties of AGN hosts changes with \Lx, we split our sample of X-ray selected AGNs in two bins of \Lx\ separated at \Lx$=2\times10^{43}$\ergps, which we refer to as the low and high \Lx\ samples. This cut was chosen to return similar numbers of AGNs between the low and high \Lx\ samples. Overall, our low \Lx\ sample contains 271 sources (50~percent of the full sample), of which 206 (76~percent of the low \Lx\ sample) have SFR upper limits and 189 (70~percent of the low \Lx\ sample) have spectroscopic redshifts, while our high \Lx\ sample contains 270 sources (50~percent of the full sample) of which 187 (69~percent of the high \Lx\ sample) have SFR upper limits and 219 (81~percent of the high \Lx\ sample) have spectroscopic redshifts.

\section{Measuring the \Rms\ distribution}
\label{sec:analysis}

\begin{figure}
 \centering
 \includegraphics[width=0.4\textwidth]{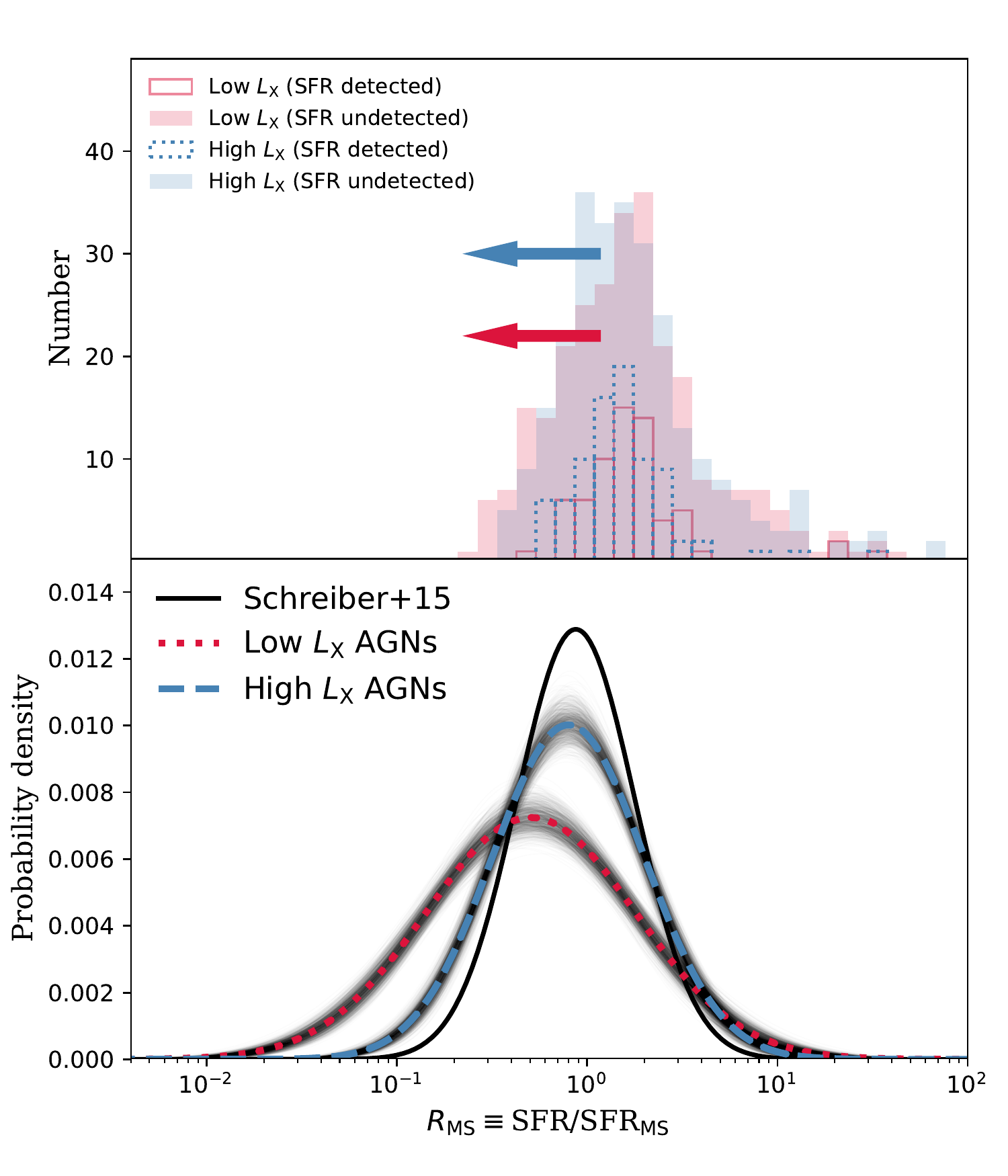}
 \caption{{\it Top}: The observed \Rms\ distributions split between low and high \Lx\ AGNs (see keys). The arrows indicate the presence of upper limits. {\it Bottom}: The optimised PDFs of \Rms\ split between low (red dotted line) and high (blue dashed line) \Lx\ AGNs. The uncertainties on the PDFs are shown with 500 black thin lines generated by randomly varying $\mu$ and $\sigma$ around their 1$\sigma$ uncertainties. We also show the \Rms\ distribution of the MS for star-forming galaxies as reported in S15. We find that the \Rms\ distribution of higher \Lx\ AGNs is narrower and closer to that of the MS of galaxies than that of the lower \Lx\ AGNs. \label{fig:distrib}}
\end{figure}

We now have a sample of 541 X-ray selected AGNs at \z=1 separated into bins of low and high \Lx\ and for which we have constraints (in terms of detections or upper limits) of SFRs and stellar masses. However, the presence of a large number of SFR upper limits ($\sim$70 percent) prevents us from directly deriving the distribution of SFRs. Instead, following M15, we assume that the distribution of SFRs relative to that of the MS (SFR/SFR$_{\rm MS}\equiv$\Rms) follows a log-normal distribution as observed for star-forming galaxies (e.g. \citealt{Sargent2012}; S15), and for which the probability density function (PDF) is defined as,

\begin{equation}
\label{eq:lognorm}
{\rm PDF} = \frac{1}{\sqrt{2\pi}\ \sigma} \times \exp\left(-\frac{(\log_{10}{(R_{\rm MS})}-\mu)^2}{2\sigma^2}\right),
\end{equation}

\noindent where $\mu$ and $\sigma$ are the mean and the standard deviation of the logarithm of \Rms, respectively. As suggested by M15, this assumption is to ease comparison between the \Rms\ distribution of AGN hosts and MS star-forming galaxies. We perform maximum likelihood estimation (MLE) to find the parameters $\mu$ and $\sigma$ that best fit the observed \Rms\ distributions for both the low and high \Lx\ sample (see Fig.\,\ref{fig:distrib} top panel). We use a MLE framework as it allows to incorporate SFR upper limits (see Grimmett et al. in prep). Due to the complexity of our likelihood function, it cannot be maximised analytically. As a consequence, we maximise it by randomly sampling the posterior distributions of $\mu$ and $\sigma$ employing the affine invariant ensemble sampler of \cite{Goodman2010} fully implemented into {\sc EMCEE}\footnote{{\sc EMCEE} is publicly available at \url{http://dfm.io/emcee/current/}} \citep{Foreman2013}. The benefit of this is that we obtain best fitting values with meaningful uncertainties that fully account for the presence of a large number of upper limits. We use flat (bounded) prior distributions and check the posterior distributions to verify that they are not constrained in any way by the choice of our prior distributions. The median value of the posterior distribution is taken as the best fit parameter, and the standard deviation as its 1$\sigma$ uncertainty.

\input{tab1.tex}

\begin{figure}
 \centering
 \includegraphics[width=0.4\textwidth]{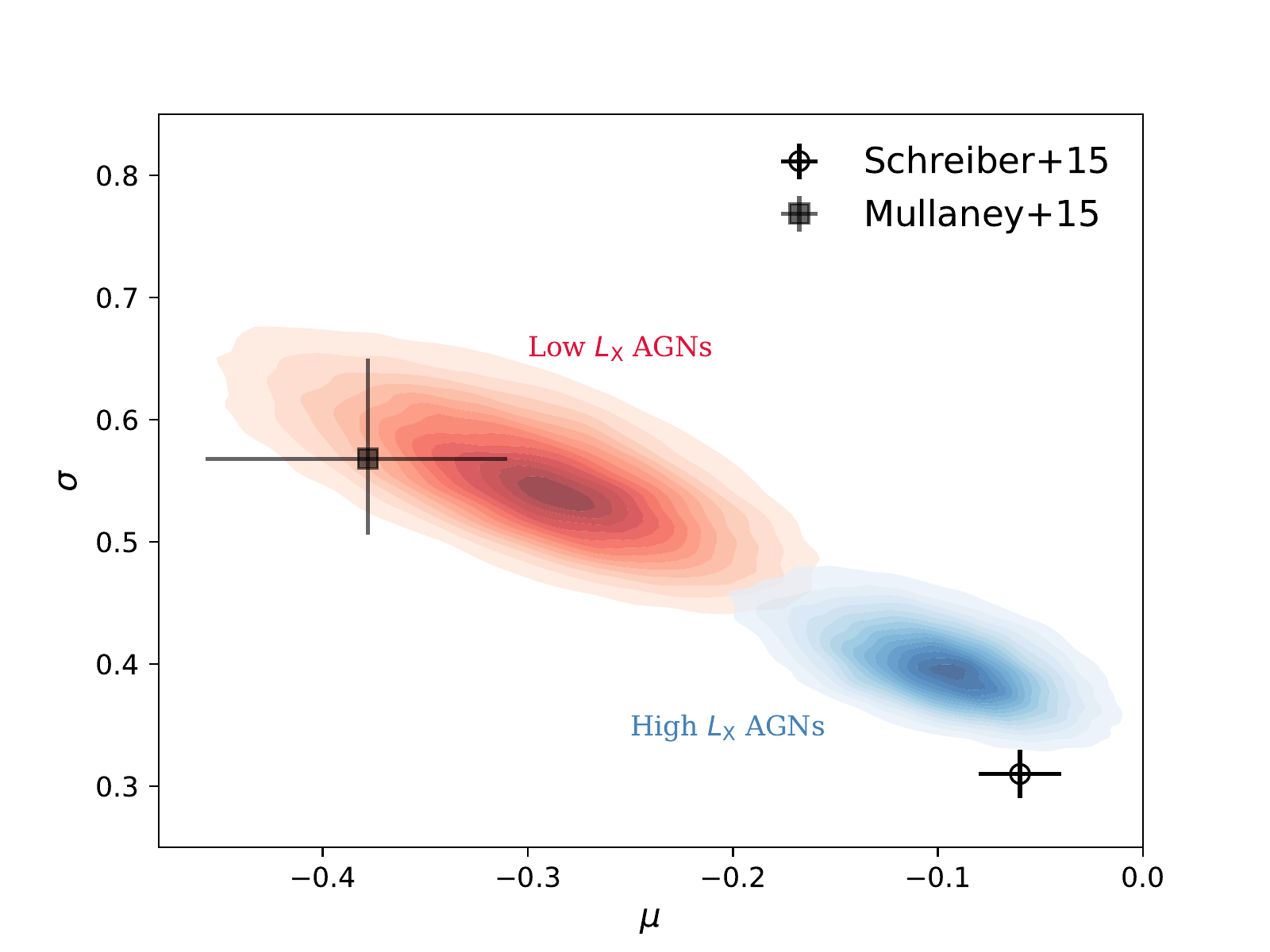}
 \caption{The bivariate distributions of $\mu$ and $\sigma$ resulting from the MLE and that define our \Rms\ distribution (see Eq.\,\ref{eq:lognorm}) at \z=1 split between low (red) and high (blue) \Lx\ AGNs. The full contours show the 1$\sigma$ spread in each case. We also show the result from M15 for AGN hosts at \z$<1.5$ and that of the MS of galaxies of S15. We find that the parameters that define the \Rms\ distribution of higher \Lx\ AGNs are more consistent with the MS than lower \Lx\ AGNs. \label{fig:densplot}}
\end{figure}

\section{Results}
\label{sec:results}

\subsection{The distributions of \Rms$\equiv$ SFR/SFR$_{\rm MS}$}
\label{subsec:rmsdistrib}

In this work we explore how the distribution of AGN host SFRs relative to that of the MS for galaxies of S15 changes with \Lx\ at \z=1. To do this, we define \Rms$\equiv$SFR/SFR$_{\rm MS}$ as the relative distance from the MS and derive its distribution assuming a log-normal shape with parameters $\mu$ and $\sigma$ (i.e. similar to that of MS galaxies) split between low and high \Lx\ AGNs (separated at \Lx=$2\times10^{43}$\ergps). Within this assumption, we find that the parameters $\mu$ and $\sigma$ for the low and high \Lx\ samples show differences (see Table\,\ref{tab:results}) and that their likelihood distributions peak at different locations in the $\mu$--$\sigma$ parameter space (see Fig.\,\ref{fig:densplot}). In particular, our results suggest that the \Rms\ distribution of higher \Lx\ AGNs is {\it narrower} (i.e. smaller $\sigma$ by a factor of 1.4) than that of lower \Lx\ AGNs (see Fig.\,\ref{fig:densplot}), indicating less diversity in the star-forming properties of higher \Lx\ AGN hosts. We also find that the \Rms\ distribution of higher \Lx\ AGN hosts peaks at a higher value of \Rms\ (i.e. higher $\mu$ by a factor of 3) than that of lower \Lx\ AGNs (see Fig.\,\ref{fig:densplot}). These can also be seen in the bottom panel of Fig.\,\ref{fig:distrib} where we show the \Rms\ distributions for low and high \Lx\ AGN hosts.

In the context of the MS for star-forming galaxies of S15, we find that, while assuming that the \Rms\ distribution of AGN hosts follows a log-normal distribution, the parameters $\mu$ and $\sigma$ of higher \Lx\ AGNs are more consistent with those reported for the MS when compared to lower \Lx\ AGNs at \z$\sim$1 (see Fig\,\ref{fig:densplot}). This suggests that the \Rms\ distribution for higher \Lx\ AGNs is in better agreement with that of MS star-forming galaxies when compared to lower \Lx\ AGNs (see Fig.\,\ref{fig:distrib} bottom panel), and that therefore higher \Lx\ AGNs at \z$\sim$1 are more likely to reside in MS star-forming galaxies than lower \Lx\ AGNs. However, interestingly, this does not prevent a large fraction of galaxies hosting lower \Lx\ AGNs from experiencing star formation at a level consistent with MS galaxies (e.g. with \Rms$>$0.4; see Fig.\,\ref{fig:distrib} bottom panel). This is a consequence of the broader \Rms\ distribution for lower \Lx\ AGNs. Finally, we also find that our results are broadly consistent with those of M15 which performed a similar analysis but for AGNs with \z$<1.5$, complemented with {\sc ALMA} data, and not split between low and high \Lx\ AGNs (see Fig.\,\ref{fig:densplot}).

\subsection{The relationship between SFR and \Lx}
\label{subsec:SFRXray}

We have explored the star-forming properties of AGNs at \z=1 by measuring the \Rms\ distribution for low and high \Lx\ AGN hosts, assuming that it has the shape of that of MS galaxies (see \S\,\ref{sec:analysis}). Our results indicate that galaxies hosting higher \Lx\ AGNs have {\it typical} SFRs consistent with MS star-forming galaxies, in contrast to galaxies hosting lower \Lx\ AGNs. 

\begin{figure}
 \centering
 \includegraphics[width=0.4\textwidth]{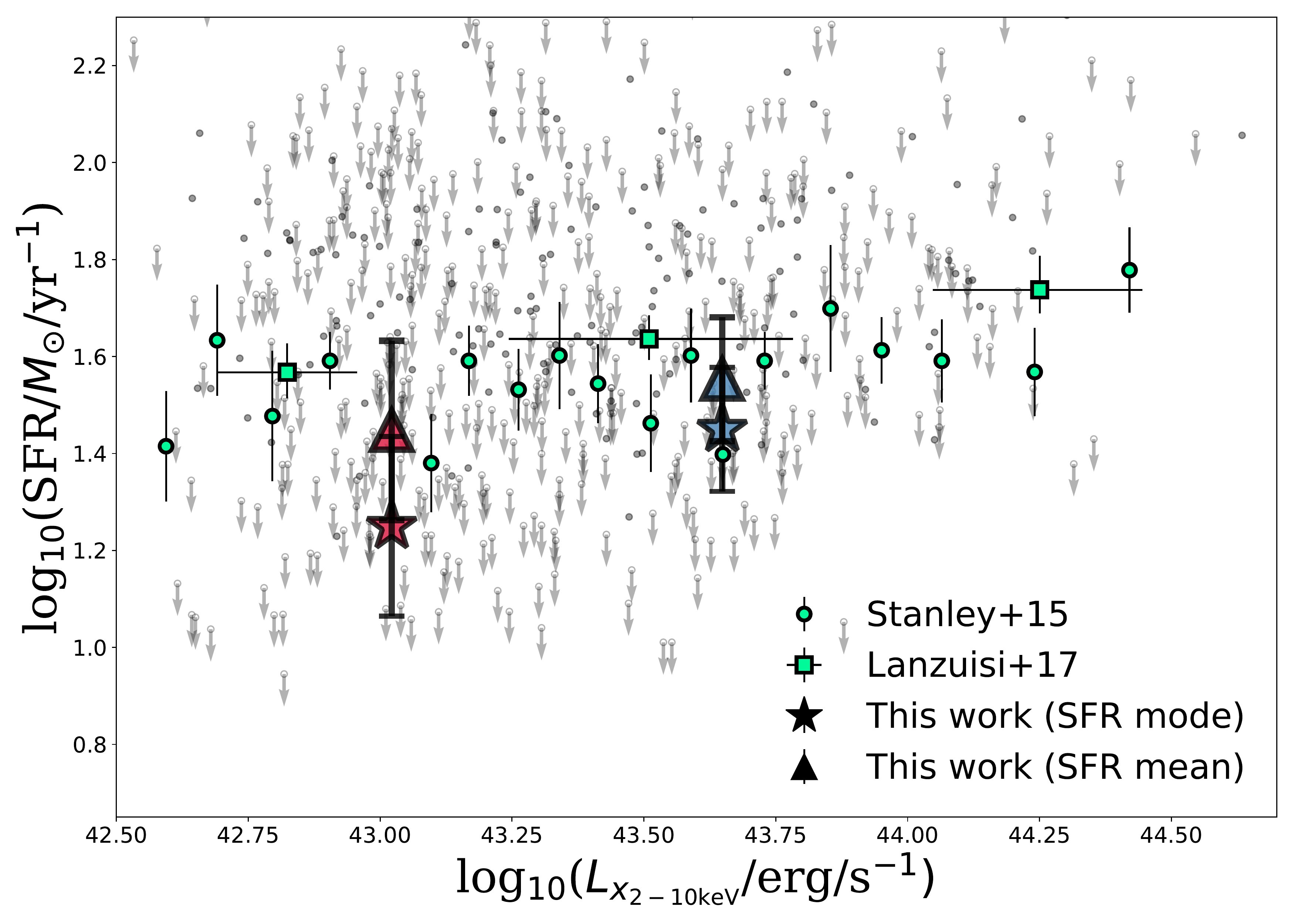}
 \caption{The relationship between SFR and \Lx\ at \z=1. The stars show our SFR modes at lower and higher \Lx. The triangles show these of our SFR means. The error bars on the mean and the mode are from propagating the uncertainties found on the parameters $\mu$ and $\sigma$ (see Table\,\ref{tab:results}). We also show the observed flat relationship found at similar redshift as reported in \protect\cite{Stanley2015} and \protect\citeauthor{Lanzuisi2017} (\protect\citeyear{Lanzuisi2017}; see bottom right-hand keys). The grey dots indicate the individual SFRs (undetected sources are shown with a downward arrow). \label{fig:SFRxray}}
\end{figure}

That the distribution of \Rms\ differs between lower and higher \Lx\ AGNs apparently contradicts recent findings of a flat relationship between SFR and \Lx\ \citep[e.g.][]{Stanley2015, Lanzuisi2017}. We investigate this apparent contradiction by measuring both the mean\footnote{The mean is defined as $R_{\rm MS}^{\rm mean} = \exp\left(\mu + \sigma^2/2\right)$.} $R_{\rm MS}^{\rm mean}$ and the mode\footnote{The mode is defined as, $R_{\rm MS}^{\rm mode} = \exp\left(\mu - \sigma^2 \right)$.} $R_{\rm MS}^{\rm mode}$ of the \Rms\ distributions for low and high \Lx\ AGNs using our best fit of $\mu$ and $\sigma$ (see Table\,\ref{tab:results}). Knowing SFR$_{\rm MS}$ for each host, we are able to derive the mean and the mode SFR for low and high \Lx\ AGNs. We show in Fig.\,\ref{fig:SFRxray} that our average SFRs are in agreement with recent studies that find a flat relationship between SFR and \Lx\ \citep[e.g.][]{Stanley2015, Lanzuisi2017}. However, as expected, our mode SFRs systematically lie below our mean SFRs, since the means are affected by bright outliers. The reason why our mean SFRs are in better agreement with the flat relationship reported by, e.g. \cite{Stanley2015}, is that they use stacking analysis to account for undetected sources which is, in essence, on-image mean-averaging. We further find that the difference between our mode and mean SFRs changes with \Lx. This is a consequence of the broader \Rms\ distribution of galaxies hosting lower \Lx\ AGNs. We stress that the differences between the mean and the mode SFR of our \Lx\ bins are consistent within the 1$\sigma$ error bars derived from propagating the uncertainties found in the parameters $\mu$ and $\sigma$ that define our \Rms\ distribution (see Table\,\ref{tab:results}).

\section{Discussion}
\label{sec:discussion}

In this work, we investigate the \Rms$\equiv$SFR/SFR$_{\rm MS}$ distribution at \z=1 of AGNs split in two bins of \Lx\ separated at \Lx=$2\times10^{43}$\ergps, and assuming that it follows a log-normal distribution, as found for that of MS star-forming galaxies (e.g. S15). Our results hint that the \Rms\ distribution for low and high \Lx\ AGNs are different. In particular, we suggest that the \Rms\ distribution of higher \Lx\ AGN hosts is narrower than that of lower \Lx\ AGNs, which is equivalent to there being less diversity in the host star-forming properties of higher \Lx\ AGNs (see \S\,\ref{subsec:rmsdistrib}). We propose that the diversity of SFRs in lower \Lx\ AGNs is a consequence of the relative ease for a galaxy to trigger a lower luminosity AGN, as suggested by the large relative number of low \Lx\ AGNs as opposed to high \Lx\ AGNs in the X-ray luminosity functions \citep[e.g.][]{Aird2017}. In addition, within our assumptions, the \Rms\ distribution of higher \Lx\ AGNs is in better agreement with that of MS galaxies. This could indicate the necessity of a significant amount of gas (i.e. enough to sustain MS host SFRs) to trigger luminous X-ray AGNs.

Our sample contains a large number of SFR upper limits. Yet, our MLE approach fully accounts for this effect by providing sensible uncertainties, and shows that, with the current level of SFR detections at \z$\sim$1 using far-infrared wavelength, it is likely that lower and higher \Lx\ AGNs have different \Rms\ distributions (i.e. see Fig.\,\ref{fig:densplot}). If confirmed, the larger agreement between the \Rms\ distribution for higher \Lx\ AGNs with that of the MS of star-forming galaxies suggests a stronger link between SMBH and galaxy growth for powerful AGNs. That is, the concurrence of a higher \Lx\ AGN -- indicating ongoing SMBH growth -- with a less diverse, more MS-like star-forming host galaxy -- suggesting host galaxy growth. The similar galaxy mass distributions between our low and high \Lx\ samples (see bottom panel of Fig.\,\ref{fig:distribs}) allows us to make such a direct link between average \Lx\ and average SMBH accretion rate by using the specific \Lx\ (i.e. \Lx/\Mstar) as a proxy for Eddington ratio (see \citealt{Bernhard2018} and references therein). In this context, our results are consistent with recent studies that find that the SMBH accretion rate changes with the host galaxy properties (e.g. \citealt{Kauffmann2009, Georgakakis2014, Wang2017, Aird2018a, Aird2018b, Bernhard2018}; Grimmett et al. subm.). Furthermore, the finding of a \Rms\ distribution for higher \Lx\ AGNs in better agreement with MS galaxies is also consistent with results showing that Quasar-like AGN activity is often found in star-forming galaxies \citep[e.g.][]{Kalfountzou2014, Rosario2013b, Stanley2017}.

Finally, we note that our results hold (i.e. narrower distribution for higher \Lx\ AGNs) when only considering galaxies with stellar masses \Mstar$>10^{10}$ \Msun\ as is often used to avoid incompleteness (e.g. \citealt{Scholtz2018}). 
 
\section{Conclusion}
\label{sec:conclusion}

We measure the \Rms\ distribution of \z=1 AGN hosts split between low and high \Lx\ AGNs separated at \Lx=$2\times10^{43}$\ergps. We use a sample of 541 X-ray selected AGNs from the {\sc COSMOS} field for which we derive SFRs or upper limits and measure stellar masses (see \S\,\ref{sec:sample}). We perform MLE to infer the \Rms\ distribution of the two samples of AGN hosts incorporating upper limits and under the assumption that the \Rms\ distribution is parametrised as a log-normal distribution, identical to that of the MS of star-forming galaxies (see \S\,\ref{sec:analysis}). Our main results show that, with this assumption, the \Rms\ distribution of higher \Lx\ AGNs is narrower (i.e. smaller $\sigma$ by a factor of 1.4) and peaks at a higher value of \Rms\ (i.e. higher $\mu$ by a factor of 3) than that of lower \Lx\ AGNs (see Fig.\,\ref{fig:densplot}). This suggests less diversity in the star-forming properties of higher \Lx\ AGNs when compared to their lower \Lx\ counterpart. We speculate that the larger diversity in the star-forming properties of lower \Lx\ AGNs may arise from the relative ease of a SMBH to trigger a low \Lx\ AGN in comparison to triggering a higher \Lx\ AGN. Furthermore, higher \Lx\ AGNs have hosts with star-forming properties in better agreement with that of MS star-forming galaxies, indicating that higher \Lx\ AGNs are more likely to reside in MS star-forming galaxies. We also investigate the relationship between SFR and \Lx\ for our two distributions by measuring the change in the mean and the mode of SFR with \Lx\ (see \S\,\ref{subsec:SFRXray}). We find that our mean and mode SFRs are consistent with the flat relationship found between SFR and \Lx, and that the mode SFRs lie below that of the mean, with a larger difference between the mean and the mode at lower \Lx (see Fig.\,\ref{fig:SFRxray}). This is a consequence of the differences in the width of the distributions at low and high \Lx\ with the mean SFR being affected at different levels by bright outliers in the low and high \Lx\ sample.

\section*{Acknowledgements}
We thank the anonymous referee. EB, JM, CT acknowledge STFC grant R/151397-11-1. EB thanks C.M. Harrison and J. Scholtz for useful discussions that help improving the clarity of the results and discussion.





\bibliographystyle{mnras}
\bibliography{./biblio}








\bsp	
\label{lastpage}
\end{document}

%% file: tab1.tex
\begin{table}
\begin{threeparttable}
\caption{The results of the MLE performed to find $\mu$ and $\sigma$ that best fit the observed \Rms\ distributions for our low and high \Lx\ sample of AGNs. Associated errors on each parameter are the 1$\sigma$ uncertainties measured from the posterior distributions (see \S\,\ref{sec:analysis}). We also show results for the AGN sample at \z$<1.5$ of \protect\cite{Mullaney2015} and for the MS of star-forming galaxies of \protect\cite{Schreiber2015}. \label{tab:results}}
\centering
\begin{tabular}{ccc}
\multirow{2}{*}{Sample}   & $\mu$ & $\sigma$ \\ \vspace{0.0cm}
   &  mean of ln(\Rms) & std. dev. of ln(\Rms) \\
\hline
This Work & \multirow{2}{*}{$-0.30^{\pm 0.06}$} & \multirow{2}{*}{$0.55^{\pm 0.05}$} \\ \vspace{0.1cm}
Low \Lx\ AGNs &  & \\

This Work & \multirow{2}{*}{$-0.10^{\pm 0.04}$} & \multirow{2}{*}{$0.40^{\pm 0.03}$}  \\ \vspace{0.1cm}
High \Lx\ AGNs &  & \\

All AGNs (\z$<1.5$) & \multirow{2}{*}{$-0.38^{+0.07}_{-0.08}$} & \multirow{2}{*}{$0.6^{\pm 0.1}$}  \\ \vspace{0.1cm}
\citep{Mullaney2015} &  & \\

Main Sequence & \multirow{2}{*}{$-0.06^{\pm 0.02}$} & \multirow{2}{*}{$0.31^{\pm 0.02}$}  \\ \vspace{0.1cm}
\citep{Schreiber2015} &  & \\

\end{tabular}%
\end{threeparttable}
\end{table}